\documentclass[preprint,prd,aps,tightenlines]{revtex4}

\usepackage{amsmath}
\usepackage{verbatim}
\usepackage{graphicx}
\usepackage[usenames]{color}
\usepackage{psfrag}

\def\beq{\begin{equation}}
\def\eeq{\end{equation}}
\def\bea{\setlength\arraycolsep{1.4pt}\begin{eqnarray}}
\def\eea{\end{eqnarray}}
\def\bit{\begin{itemize}}
\def\eit{\end{itemize}}

\begin{document}

\title{Gravity heats the Universe}

\author{Adam Moss} \email{adammoss@phas.ubc.ca}
\affiliation{Department of Physics \& Astronomy\\
University of British Columbia,
Vancouver, BC, V6T 1Z1  Canada}
\author{Douglas Scott} \email{dscott@astro.ubc.ca}
\affiliation{Department of Physics \& Astronomy\\
University of British Columbia,
Vancouver, BC, V6T 1Z1  Canada}

\date{\today}

\begin{abstract}
Structure in the Universe grew through gravitational instability from very
smooth initial conditions.  Energy conservation requires that the growing
negative potential energy of these structures is balanced by an increase in
kinetic energy.  A fraction of this is converted into heat in the collisional
gas of the intergalactic medium.
Using a toy model of gravitational heating we attempt to link the 
growth of structure in the Universe and the average temperature of this
gas.  We find that the gas is rapidly heated 
from collapsing structures at around $z\,{\sim}\,10$, reaching a temperature 
${>}\,10^6$K today, depending on some 
assumptions of our simplified model.  Before that there was a cold era from 
$z\sim100$ to $\sim10$ in which the matter temperature is below that of 
the Cosmic Microwave Background.
\end{abstract}

\maketitle
\noindent
{\em Introduction.---}%
It is well known that the temperature of the Cosmic Microwave 
Background is $T_{\gamma 0}=2.725\,$K, while the atomic matter today is 
typically orders of magnitude hotter than this.  As the Universe 
expands the photon temperature drops so that $T(z)=T_{\gamma 0}(1+z)$, 
and Compton scattering couples the matter to the radiation until about 
$z\,{\simeq}\,100$, when it is allowed to cool adiabatically with 
$T\propto(1+z)^2$.  In a perfectly smooth Universe with the same 
background cosmology as we observe today, the present temperature of 
the gas would be only about $20\,$mK~\cite{Scott:2009sz}.

However, the Intergalactic Medium (IGM) is composed of gas at a variety 
of temperatures (e.g.~\cite{Valageas:2001wh}), ranging from 
${\sim}\,10^4$K in dense clouds, up to ${>}\,10^6$K in the rarest 
inter-cloud gas.  The reason that the gas is so hot is patently because 
the Universe contains structure.  Our lumpy Universe generates thermal energy
in a number of ways, but the simplest source is purely gravitational.
The importance of gravitational heating for understanding the structure and 
evolution of  galaxy clusters has been discussed by a number of authors (e.g.\
see~\cite{Dekel:2007zy,Khochfar:2007rp} for recent examples). However, 
an {\it explicit\/} connection between gravity and heating of the IGM 
seems to have escaped notice in the literature. The gas is effectively 
heated from the growth of structure through gravitational instability, 
with the thermal energy coming from the increasingly negative energy of the 
growing potential  wells.

The idea of gravitational shock heating of the IGM as a direct result 
of structure formation goes back at least to the `pancake' model of the 
1970s~\cite{sz}. This work was later extended using the `Zel'dovich 
approximation' to follow the shock-heating of gas outside collapsed 
objects (see e.g.~\cite{Nath:2001yd}), or in a related approach to
use an extension of 
the Press-Schechter formalism to estimate the 
fraction of shocked gas~\cite{Furlanetto:2003vk}.
Such numerical calculations allow for an investigation of the contributions
of collapsed and shocked gas to temperature evolution of the IGM
(see also \cite{Pen:1998eg,Springel:2000bq,Wang:2008zz}).

The details will of course be quite complicated, not least because starbursts
and quasars provide photon and mechanical heating to the IGM in very
non-linear and inhomogeneous processes.  However, we will here focus only
on gravity 
and aim to estimate the temperature of the IGM from the 
potential energy of the Universe, using a toy model to 
highlight the basic connection between gravitational and thermal 
energies. 
For our numerical work, 
we use the baryon density $\Omega_{\rm b} = 0.045$, cold dark matter 
density $\Omega_{\rm c} = 0.0255$, Hubble parameter $H_0= 70  \rm 
\,{\rm km} \, {\rm sec}^{-1} \, {\rm Mpc}^{-1} $, and the spectral 
index of initial fluctuations $n_{\rm s}=1$. To normalize the power 
spectrum we fix the matter variance in $8 \, h^{-1} \, {\rm Mpc}$ 
spheres to $\sigma_8=0.9$. These values are consistent with the current 
best fit cosmological parameters~\cite{Dunkley:2008ie}.

\begin{figure}
\centering
\mbox{\resizebox{0.55\textwidth}{!}{\includegraphics[angle=0]{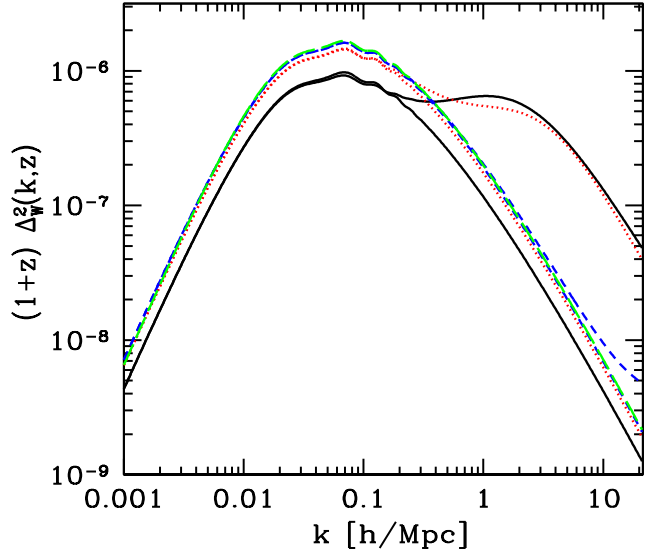}}}
\caption{\label{fig:w_pe} Normalized gravitational potential as a 
function of scale at $z=0$ (solid), 1 (dotted), 10 (short-dashed) and 
100 (long-dashed). The corresponding curves with additional
small-scale power are non-linear corrections from the {\tt HALOFIT}
code~\cite{Smith}.}
\end{figure}

\noindent
{\em Potential energy estimate from power spectrum.---}%
The volume averaged gravitational potential energy (GPE) per unit mass 
as a function of redshift $z$ is~\cite{Siegel:2005xu}
\beq
W (z) =\frac{1}{2} \langle (1+\delta({\bf x},z) ) \phi({\bf x},z)  
\rangle\,,
\eeq
where $\phi$ is the Newtonian potential, defined by the line element 
$ds^2= a^2 \left[ -(1+2\phi) d\tau +(1-2\phi) dx^2 \right] $, and 
$\tau$ is conformal time. Using the Poisson equation $\nabla^2 \phi = 
4\pi G a^2 \rho \, \delta$ and the definition of the correlation 
function 
\beq
\xi({\bf r}, z) \equiv \langle \delta ({\bf x}, z) \delta ({\bf x}+{\bf 
r}, z)   \rangle =  \frac{1}{(2\pi)^3} \int d^{3} {\bf k} \, { P}(k, z) 
e^{i \bf{k \cdot r}}\,,
\eeq
where $P(k, z) \equiv |\delta(k, z)|^2$, the Newtonian potential energy 
is
\beq
W(z)=- \int \frac{dk}{k} \Delta_{\rm W}^2 (k,z)\,,
\eeq
and with the dimensionless gravitational power
\beq
\Delta_{\rm W}^2 (k,z) \equiv \frac{3 H_0^2 \Omega_{\rm m, 0} (1+z)}{8 
\pi^2} P(k,z) \, k\,.
\eeq
This can also be obtained from the definition in 
Peebles~\cite{Peebles}, $W=-\frac{1}{2} G \rho_{\rm m} a^2 \int d^3 
{\bf r}  \, \xi({\bf r}, z)/{\bf r}=- 2\pi G \rho_{\rm m} \, J_2$, 
where $J_2= \int dk P(k, z)/(2\pi^2)$.  In  Fig.~\ref{fig:w_pe} we show 
$(1+z) \Delta_{\rm W}^2(k,z)$ for our cosmological model at $z=0, 
1, 10$ and $100$. The peak of the gravitational energy contribution occurs at a 
comoving scale of $\sim 10$ Mpc. 

In the matter dominated era ($z \gtrsim 1$) the quantity $(1+z) \Delta_{\rm 
W}^2(k,z)$  remains roughly constant, since the growth factor, given by 
$D(z)=\delta_{\rm m} (z)/\delta_{\rm m} (z=0)$ , scales as 
$(1+z)^{-1}$. At lower $z$, in the dark energy dominated era, the 
growth of structure is significantly suppressed due to the increasing 
expansion rate -- this slow down has recently been detected in 
observations of galaxy clusters~\cite{Vikhlinin:2008ym}.

\begin{figure}
\centering
\mbox{\resizebox{0.55\textwidth}{!}{\includegraphics[angle=0]{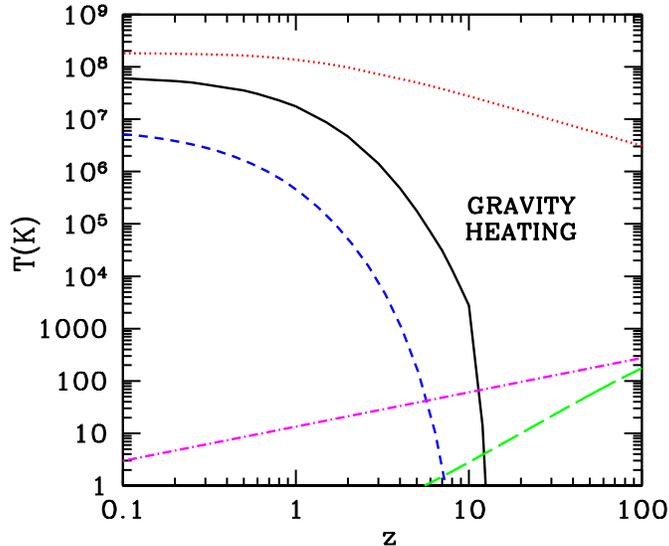}}}
\caption{\label{fig:tz} Evolution of $T_{\rm gas}(z)$ for linear theory (dotted 
curve), non-linear theory with a cut-off scale (solid) and halo number 
density (short-dashed) . We also show the baryonic temperature 
evolution from the {\tt RECFAST} code~\cite{Seager:1999km,Wong:2007ym} 
(long-dashed) and the CMB temperature (dot-dashed).} 
\end{figure}

In order to estimate the IGM temperature $T_{\rm gas}(z)$ one can suppose that
the GPE is equal to the average kinetic density energy of baryonic matter. 
Therefore, one could set $\rho_{\rm m} W = \rho_{\rm b} k T/m_{\rm H}$,
where $k$ is Boltzmann's constant, $m_{\rm H}$ the mass of 
hydrogen (neglecting helium for simplicity)
and $\rho_{\rm m}$, $\rho_{\rm b}$ are the energy densities of 
matter and baryons respectively.  We show the results of this 
calculation in Fig.~\ref{fig:tz}, along with $T(z)$ from the recombination
code {\tt RECFAST}~\cite{Seager:1999km,Wong:2007ym}, which assumes the matter 
distribution is perfectly smooth  and cools adiabatically. One finds an 
average temperature of $\sim 10^8$ K at $z=0$,  and the `gravitational 
temperature' exceeds the {\tt RECFAST} value for $z  < 1000$.

Clearly, this linear calculation over-predicts the temperature. The reason 
for this is  that gravitational shock-heating is associated with the 
collapse of {\em non-linear\/} objects and shell-crossing.
The linear GPE is largest on 
scales of $\sim 10$ Mpc, which are only {\em just\/} going non-linear at 
$z=0$. Linear scales are associated with smooth bulk flows, and 
hence there is no mechanism for the gas to be heated.  To be more realistic, we should only include the total GPE coming from
non-linear scales at a given redshift.  In this regime, setting 
$\rho_{\rm m} W = \rho_{\rm b} k T/m_{\rm H}$ implicitly assumes that the damped outgoing shock waves from collapsed 
objects efficiently (and instantaneously) heat the IGM.

 Since the linear power spectrum becomes 
becomes inaccurate on small-scales, we include non-linear corrections 
to $P(k, z) $ from the the {\tt HALOFIT} code~\cite{Smith}. These corrections 
are shown in Fig.~\ref{fig:w_pe}, and result in an increase of 
small-scale power. In order to estimate the cut-off scale at which 
perturbations are going non-linear, we compute the value $R_{\rm NL}$ 
at which the RMS mass variance 
\beq
\delta_{\rm R}^{2} (z)=\int \frac{dk}{k} \Delta^2 (k,z) W^2 (k R)\,,
\eeq 
is equal to unity. Here $W (x)$ is the window function associated with 
a spherical top-hat, and 
$\Delta^2 (k,z)=  k^3 P(k,z)/(2\pi^2)$ is the dimensionless power 
spectrum.  We then perform the integral only above $k_{\rm NL}=2\pi/R_{\rm 
NL}$. The results of this computation are shown in Fig.~\ref{fig:tz} -- 
one finds a smaller temperature at $z=0$, and a much faster decrease 
with redshift. This is due to $k_{\rm NL}$ shifting to smaller-scales 
for increasing $z$, so less GPE contributes to the heating.  


\noindent
{\em Halo number density estimate.---}%
We can also estimate the gravitational energy in a different way by 
just considering viralized objects. Virialization occurs at higher 
over-densities than those discussed previously -- in a flat 
Einstein-de-Sitter model linear theory predicts a spherical top-hat 
will break away from the expansion at a mass fluctuation of 
$\delta=1.063$, collapse at $\delta=1.686$, but virialize at  
$\delta=1.59$ (see e.g.~\cite{padmanabhan}). We use the Press-Schechter 
mass function~\cite{ps} to estimate the number density of viralized 
objects
\beq
\frac{dn}{dM} (z, M) = \sqrt{\frac{2}{\pi}} \frac{\rho_{\rm m} 
(t_0)}{M^2} \nu e^{-\nu^2/2} \left[ -\frac{d \ln \sigma_{\rm M} (z=0) 
}{d \ln R} \right]\,,
\eeq
where $\nu\equiv\delta_{\rm c}/ \left[ D(z) \sigma_{\rm M} (z=0) \right]$, 
the critical threshold $\delta_{\rm c}=1.686$, the mass enclosing a 
sphere of radius $R$ is $M=4\pi \rho_{\rm m} (t_0) R^3/3 $ and the 
growth function is normalized to unity today.  The growth function can 
be computed from 
\beq
D(a)=\frac{5  \Omega_{\rm m, 0} }{2} \frac{H(a)}{H_0} \int^{a}_0 
\frac{da^{\prime}}{(a^{\prime} H(a^{\prime})/H_0)^3}\,,
\eeq
where the scale-factor $a=1/(1+z)$ and $H(a)$ is the Hubble rate.

In order to compute the gravitational energy we can use the virial 
theorem, $W=-2K$, where the kinetic energy per unit volume is $K= k 
\langle T_{\rm vir} \rangle M_{\rm tot}$. The mass averaged virial 
temperature is defined as
\beq
\langle T_{\rm vir} \rangle = \frac{1}{M_{\rm tot}} \int^{\infty}_0 
T_{\rm vir} M \frac{dn}{dM} dM 
\eeq
with $M_{\rm tot}= \int^{\infty}_0 M \frac{dn}{dM} dM$. We use the 
virial temperature normalization for a conventional cosmology 
in~\cite{Pierpaoli:2000ip}, with $T_{\rm vir} ({\rm K})=5 \times 
10^{-23} \, (M/{\rm kg})^{2/3} (1+z)$.

We again assume that outgoing shock waves effectively share out the GPE among
all the particles in the Universe, not just those in virialized structures.
Hence one can equate
$k \langle T_{\rm vir} \rangle M_{\rm tot}=\rho_{\rm b} k T$
to obtain the evolution of $T(z)$. This is shown 
in Fig.~\ref{fig:tz} -- the redshift evolution has a similar profile to 
the non-linear power spectrum estimate (with a cut-off scale $R_{\rm NL}$),
although the temperature is approximately an order of magnitude 
lower.  One can see that effectively these two calculations are very similar,
with mainly just a different choice of cut-off scale.

\begin{figure}
\centering
\mbox{\resizebox{0.55\textwidth}{!}{\includegraphics[angle=0]{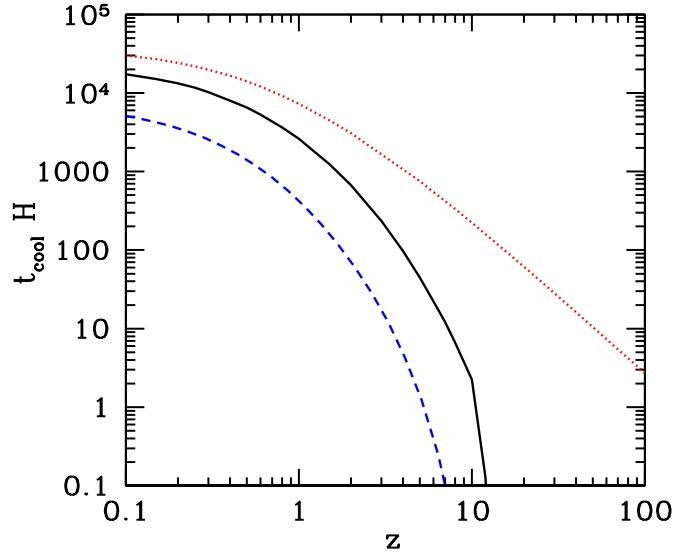}}}
\caption{\label{fig:cool} Ratio of cooling time to Hubble time. 
Lines are labelled the same as in Fig.~\ref{fig:tz}.}
\end{figure}

\noindent
{\em Gas cooling.---}%
So far, we have assumed that the IGM is heated without taking into 
account cooling processes in the gas. We can expect this to be a 
reasonable estimate as long as the cooling time-scale is longer than 
the  Hubble time $H^{-1}$. Assuming the cooling is dominated by thermal 
bremsstrahlung we show the ratio of cooling to Hubble time in 
Fig.~\ref{fig:cool}. For our model of IGM heating from only collapsed 
objects, we find the cooling time becomes less than the Hubble time 
around $z\,{\sim}\,10$. At this point the baryonic matter in the IGM will 
begin to heat from its adiabatically cooled value of ${\sim}\,3\,$K.  The 
CMB temperature is ${\sim}\,30\,$K at $z\,{\sim}\,10$, so it appears there is 
an epoch from $z\,{\sim}\,10$--100 when the matter is actually colder.  
Since the IGM is outside collapsed regions and thus taking part in the
Hubble expansion, it will also continue to cool 
adiabatically -- including this additional effect in our computations 
leads to a reduction in temperature by a factor of 2--3 at $z=0$.

\noindent
{\em Conclusions.---}%
Since the Universe is lumpy, it is necessarily hot.  A simplistic picture is
that the temperature of the IGM comes from energy balance with a fraction of
the gravitational energy that is building up through gravitational instability.
Virliaization of structures require that about half of the potential energy
is lost to the `environment', here meaning that the dark and baryonic matter
acquire kinetic energy.  When shell-crossing occurs kinetic energy is
converted into thermal energy in the collisional material, the IGM gas.
We have shown that this picture leads to an IGM which is significantly colder
than the CMB until gravitational heating takes over at $z\,{\sim}\,10$.

Things became rapidly more complicated just after this era.
An accurate calculation
of the heating process would require solving for the inhomogeneous growth
of structure, including hydrodynamic effects, as well as cooling processes.
In addition, reionization of the Universe appears also to happen at
$z\,{\sim}\,10$ and so photon sources, electromagnetic interactions and
radiative transfer need to be considered in order to fully understand the IGM
today.  Many astrophysics theorists are working hard on just these problems.

{\em Acknowledgments.---}%
This research was supported by the Natural Sciences and Engineering 
Research Council of Canada.  We thank the many colleagues with whom we have
had fruitful discussions on this topic, in particular Alan Duffy, Kris
Sigurdson, Martin White and Jim Zibin.


\begin{thebibliography}{99}
\newcommand{\prlet}{Phys.\ Rev.\ Lett.}
\newcommand{\npb}{Nucl.\ Phys.\ B}
\newcommand{\pletb}{Phys.\ Lett.\ B}
\newcommand{\prevd}{Phys.\ Rev.\ D}
\newcommand{\jhep}{J.\ High Energy Phys.}
\newcommand{\cqg}{Class.\ Quant.\ Grav.}
\newcommand{\jast}{Astrophys. \ J.}

\bibitem{Scott:2009sz}   
D. Scott and A. Moss, (2009) [arXiv:0902.3438].

\bibitem{Valageas:2001wh}
P. Valageas, R. Schaeffer and J. Silk, {\em Astron. Astrophys.} {\bf 
388}, (2002) 741 [astro-ph/0112273].

\bibitem{Dekel:2007zy}
A. Dekel and Y. Birnboim, {\em MNRAS} {\bf 383}, (2008) 119 
[arXiv:0707.1214]

\bibitem{Khochfar:2007rp}
S. Khochfar and J. P. Ostriker, {\em \jast} {\bf 680}, (2008) 54 
[arXiv:0704.2418].

\bibitem{sz}
R. A. Sunyaev and Y. B. Zel'dovich, {\rm Astron. Astrophys.} {\bf 20}, 
(1972) 189.

\bibitem{Nath:2001yd}
B. B. Nath and J. Silk, {\em MNRAS} {\bf 327}, (2001) 5 
[astro-ph/0107394].

\bibitem{Furlanetto:2003vk}
S. Furlanetto and A. Loeb, {\em \jast} {\bf 611}, (2004) 642 
[astro-ph/0312435].

\bibitem{Pen:1998eg}
U. Pen, {\em \jast} {\bf 510}, (1999) 1 [astro-ph/9811045].

\bibitem{Wang:2008zz}
P. Wang and T. Abel, {\em \jast} {\bf 672}, (2008) 752.

\bibitem{Springel:2000bq}
V. Springel, M. White and L. Hernquist, (2000) [astro-ph/0008133].

\bibitem{Dunkley:2008ie}   
J. Dunkley {\em et al}, {\em Astrophys. \ J.\.Supp.} {\bf 180}, (2009) 
306 [arXiv:0803.0586].

\bibitem{Siegel:2005xu}
E. R. Siegel and J. N. Fry, {\em \jast} {\bf 628}, (2005) 1 
[astro-ph/0504421]. 

\bibitem{Peebles}
P. J. E. Peebles, {\em Physical Cosmology}, Princeton University Press 
(1971) .

\bibitem{Smith}
R. E. Smith {\em et al}, {\em MNRAS} {\bf 341}, (2003) 1311 
[astro-ph/0207664].

\bibitem{Vikhlinin:2008ym}   
A. Vikhlinin {\em et al}, (2008) [arXiv:0812.270].

\bibitem{Seager:1999km}
S. Seager, D. D. Sasselov and D. Scott, {\em \em Astrophys. \ 
J.\.Supp.} {\bf 128}, (2000) 407 [astro-ph/9912182].

\bibitem{Wong:2007ym}
W. Y. Wong, A. Moss and D. Scott, {\em MNRAS} {\bf 386}, (2008) 1023 
[arXiv:0711.1357].

\bibitem{padmanabhan}
T. Padmanabhan, {\em Theoretical Astrophysics: Galaxies and Cosmology}, 
Cambridge University Press (2002) .

\bibitem{ps}
W. H. Press and P. Schechter, {\em \jast} {\bf 187}, (1974 452. 

\bibitem{Pierpaoli:2000ip}
E. Pierpaoi, D. Scott and M. White, {\em MNRAS} {\bf 325}, (2001) 77 
[astro-ph/0010039].


\end{thebibliography}
\end{document}